\documentclass[%
 reprint,
 amsmath,amssymb,
 aps,
]{revtex4-1}

\usepackage{graphicx}% Include figure files
\usepackage{dcolumn}% Align table columns on decimal point
\usepackage{bm}% bold math
\usepackage{float}
\begin{document}
\preprint{APS/123-QED}

\title{Thermal Conductivity Coefficient from Microscopic Models}

\author{T.E Nemakhavhani}
 \email{200824451@student.uj.ac.za}
\affiliation{%
 University of Johannesburg,South Africa\\
}%
\author{A. Muronga}
\affiliation{
 University of Johannesburg,South Africa\\
}%
%\affiliation{
% iThemba LABS
%}%
%\author{S. M. Mullins, A. E. Lawrie, J. J. Lawrie, R. A. Bark, P. Papka, P. L. Masiteng}
%\affiliation{%
 %iThemba LABS\\
%}%

\date{\today}% It is always \today, today,

\begin{abstract}

Thermal conductivity of hadron matter is studied using a microscopic transport model, which will be used to simulate ultra-relativistic heavy ion collisions at different energy densities $\varepsilon$,
%can support the newly LHC energy of up to $\sqrt{s} =$ 14 TeV
namely the Ultra-relativistic Quantum Molecular Dynamics (UrQMD). The molecular dynamics simulation is performed for a system of light mesons species $(\pi, \rho, K)$ in a box with periodic boundary conditions. Equilibrium state is investigated by studying chemical equilibrium and thermal equilibrium of the system. Particle multiplicity equilibrates with time, and the energy spectra of different light mesons species have the same slopes and common temperatures when thermal equilibrium is reached. Thermal conductivity transport coefficient is calculated from the heat current - current correlations using the Green-Kubo relations. 
\end{abstract}

\maketitle

%\tableofcontents

\section{\label{sec:level1}INTRODUCTION} 
 A large number of studies in heavy ion physics and high energy physics have been done using the results from Relativistic Heavy Ion Collider (RHIC). Now with the restart of the Large Hadron Collider (LHC) physics programme, the field of high energy nuclear physics, and especially heavy ion physics, has gone into a new era. It is now possible to explore the properties of Quantum-Chromo-Dynamics (QCD) at unprecedented particle densities and temperatures at much higher energies than that produced at RHIC from $\sqrt{s} =$ 200 GeV to $\sqrt{s} =$ 14 TeV at LHC \cite{steinheimer2013nonthermal,campbell2011vector}.

High energy heavy ion reactions are studied experimentally and theoretically to obtain information about the properties of nuclear matter under extreme conditions at high
densities and temperatures as well as about the phase transition to new state of matter, the quark-gluon plasma (QGP) \cite{bratkovskaya2000aspects, demir2014hadronic,muronga2004shear}.
This work reports on the transport coefficient namely thermal conductivity of hadron matter. Other transport coefficients such as shear and bulk viscosity are well discussed and
documented \cite{gorenstein2008viscosity, muronga2004shear}, but the study of thermal conductivity transport coefficient is poorly documented especially with the use of UrQMD model to simulate ultra-relativistic heavy ion collisions. The knowledge of this transport coefficient plays an important role in the development of the model such as UrQMD model and also the development of high energy heavy ion experiments such as LHC and RHIC. 

To study transport coefficients of hadronic matter, an equilibrated system is considered. Hence equilibration of this system is studied in an infinite hadronic matter using microscopic transport model which can now support new LHC energy of $\sqrt{s}=$ 2.76 TeV for lead+lead (Pb+Pb) central collisions the UrQMD. \cite{bratkovskaya2000aspects,bass1998microscopic,PhysRevC.79.044901,becattini2013hadron,bleicher2015recent}.

Equilibration of the system is studied by evaluating particle number densities from chemical equilibrium, energy spectra as well as the temperatures from thermal equilibrium of different light meson species in a cubic box which imposes periodic boundary conditions. This means that if a particle leaves the box another particle with the same momentum enters the box from either side \cite{song1997chemical}. 

The infinite hadron matter is modelled by initializing the system with light meson species namely the pion ($\pi$), the rho ($\rho$) and the kaon ($K$). We also pay our attention to the equation of state of hot dense hadron gases as it is an important quantity in high energy nuclear physics. The knowledge of the equation of state (EoS) is important for better understanding of the final state of interactions which is dominated by hadrons produced during ultra-relativistic heavy ion collisions. The EoS of nuclear matter is one of the key points to gain further understanding since EoS directly provides the relationship between the pressure and the energy at a given net-baryon density \cite{li2009effects}. The thermodynamic properties, transport coefficients and EoS for hadron gas are a major source of uncertainties in dissipative fluid dynamics \cite{muronga2004shear,bleicher2015recent,muronga2007relativistic}.
 
We focus on the hadronic scale temperature (100 MeV $<$ $T$ $<$ 200 MeV) and zero baryon number density which is expected to be realized in the central high energy nuclear collisions \cite{PhysRevC.79.044901}. We then change energy density from $\varepsilon$ = 0.1 - 0.5 GeV/fm$^{3}$ and for each energy density we run the system with 200 events while keeping volume and baryon number density constant until the equilibrium state is reached. Thermal conductivity transport coefficient is calculated from the heat current - current correlations using the Green-Kubo relations.

The rest of the paper is organized as follows: In section 2 we study the description of the UrQMD model. In section 3 we study equilibration properties of the system. In section 4 we study the equation of state (EoS) of the hadron gas. In section 5 we calculate thermal conductivity transport coefficient around equilibrium state through UrQMD model using Green-Kubo relations.

\section{Short description of the UrQMD model} 
The Ultra-relativistic Quantum Molecular Dynamic model (UrQMD) is a microscopic model based on a phase space description of nuclear reactions. We use the current version of UrQMD namely the UrQMD 3.3. The UrQMD 3.3 hybrid approach extends previous ansatzes to combine the hydrodynamic
and the transport models for the relativistic energies, combining these approaches into one single framework it is done for a consistent description of the dynamics \cite{steinheimer2013nonthermal}.

The UrQMD model describes the phenomenology of hadronic interactions at low and intermediate energies from few hundreds MeV up to the newly LHC energy of $\sqrt{s}$ =14 TeV per nucleon in the centre of the mass system \cite{PhysRevC.79.044901,belkacem1998equation,bleicher1999relativistic,panda2005strong}. The UrQMD collision term contains 55 different baryon species and 32 meson species which are supplemented by their corresponding anti-particles and all the isospin-projected states \cite{bass1998microscopic,bleicher1999relativistic}. The properties of the baryons and the baryon-resonances which can be populated in UrQMD can be found in \cite{bass1998microscopic} together with their respective mesons and the meson-resonances. A collision between two hadrons will occur if 
\begin{equation}
d_{trans}\le\sqrt{\frac{\sigma_{tot}}{\pi}}, \qquad \sigma_{tot}=\sigma(\sqrt{s},type).
\end{equation}
Where $d_{trans}$ and $\sigma_{tot}$ are the impact parameter and the total cross-section of the two hadrons, respectively \cite{bleicher1999relativistic}. In the UrQMD model the total cross-section $\sigma_{tot}$ depends on the isospins of colliding particles, their flavour and the centre-of-mass (c.m) energy $\sqrt{s}$. More details about the UrQMD model is presented in \cite{bass1998microscopic,bleicher1999relativistic,panda2005strong}.

\section{Equilibration of hadronic matter}
%\quad
%\newline
To investigate the equilibrition of a system the UrQMD model is used to simulate the ultra-relativistic heavy ion collisions. A multi-particle production plays an important role in the equilibration of the hadrons gas \cite{muronga2004shear}. The system of cubic box used for this study is initialized according to the following number of particle for each meson i.e 80 pions, 160 rhos and 80 kaons. For this study a cubic box with volume $V$ and a fixed baryon number density $n_{B}$ = 0.0 fm$^{-3}$ is considered. 

Thermal conductivity transport coefficient for hadron gas can be obtained by using a microscopic model that includes realistic interactions among hadrons. Thus one can adopt a relativistic microscopic model, UrQMD and perform molecular dynamical simulations for hadron gas of mesons \cite{muronga2004shear}. 

The energy density $\varepsilon$, volume $V$ and the baryon number density $n_{B}$ 
in the box are fixed as input parameters and are conserved throughout the simulation. The energy density is defined as $\varepsilon=\frac{E}{V}$, where $E$ is the energy of \textit{N} particles given by 

\begin{equation}
E=\sum_{i=1}^{N}\sqrt{m_{i}^{2}+p_{i}^{2}},
\end{equation}

and the three-momenta $p_{i}$ of the particles in the initial state are randomly distributed in the centre of mass system of the particles that is
\begin{equation}
\sum_{i = 1}^{N}p_{i} = 0.
\end{equation}

The time evolution of the particles is described by the UrQMD. We consider time evolutions of the particle densities and the energy distributions of different meson species. The system tends towards equilibrium state when the time and the energy increases as discussed in subsection 3.1 and section 3.2. 

\subsection{Chemical Equilibrium}
%\quad
%\newline
In this subsection the chemical equilibrium is studied from the particle number densities of different light meson species in a box with $V$ = 1000 fm$^{3}$, zero net baryon number density $n_{B}$ = 0.0 fm$^{-3}$ at different energy densities using UrQMD box calculations. Figure 1 and figure 2 represent the time evolution of the various meson number densities at $\varepsilon$ = 0.2 and 0.3  GeV/fm$^{3}$ energy densities. In figure 1 and figure 2 the results obtained where very small thus formula (4) is used to rescale those results for better analysis of the results and for a much readable plots
\begin{equation}
n=\frac{n_{\pi,\rho,K}\times L_{\pi,\rho,K}}{V},
\end{equation}
 
 where 
 $n_{\pi,\rho,K}$ is the particle number density (fm$^{-3}$) for each meson consideredin this UrQMD model and $L_{\pi,\rho,K}$ is the scailing factor for each meson. In this calculation $L_{\pi}$ = 1 fm$^{3}$, $L_{\rho}$ = 200 fm$^{3}$ and $L_{K}$ = 100 fm$^{3}$.

\begin{figure}[H]
\includegraphics[width=0.35\textwidth]{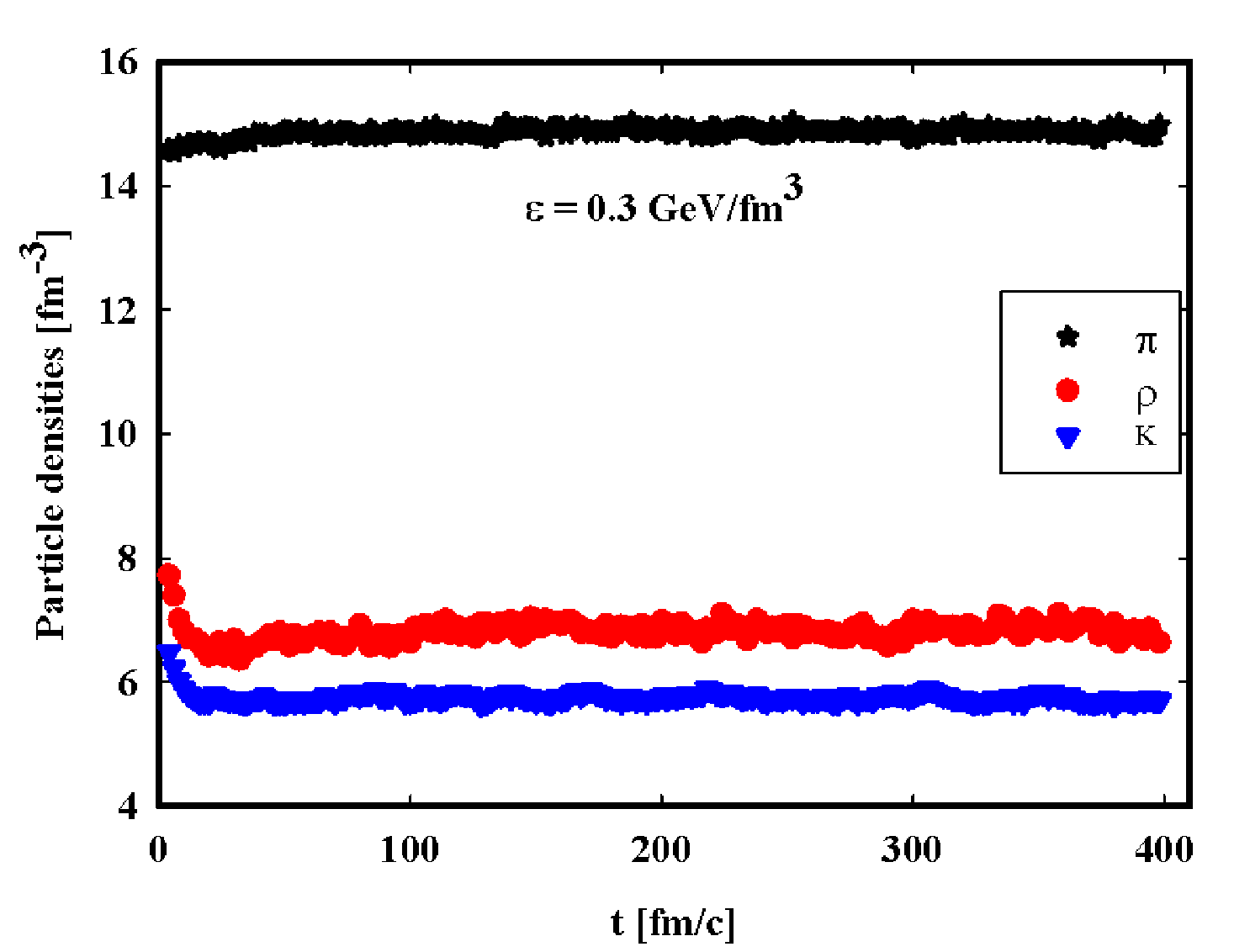}
\caption{The time evolution of particle number densities of light meson species, ($\pi$, $\rho$ and K) at $\varepsilon = 0.3$ GeV/fm$^{3}$.}
\label{N1}
\end{figure}

\begin{figure}[H]
\includegraphics[width=0.35\textwidth]{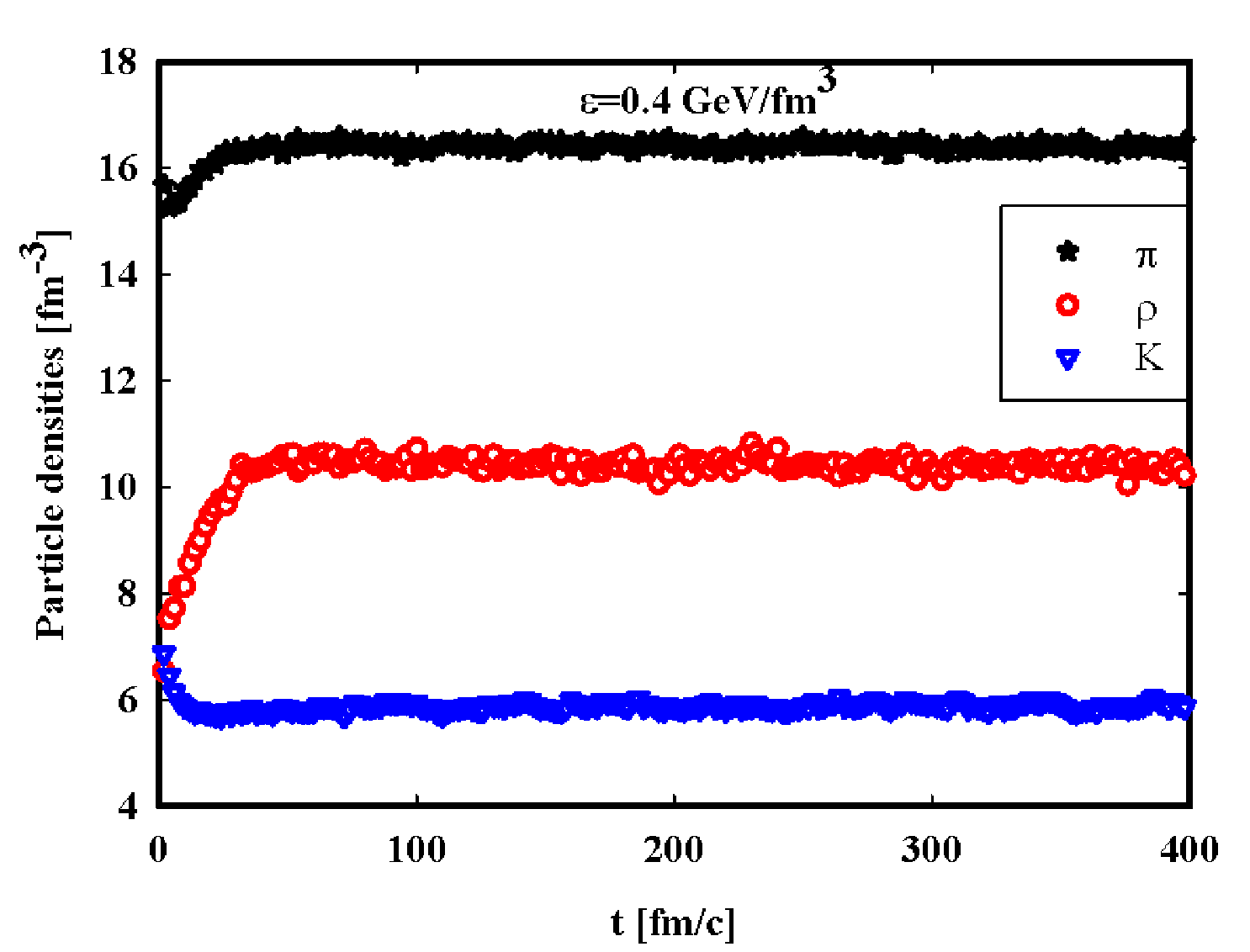}
\caption{The time evolution of particle number densities of light meson species, ($\pi$, $\rho$ and K) at $\varepsilon = 0.4$ GeV/fm$^{3}$.}
\label{D2}
\end{figure}

%\newpage
In the figures 1 and 2 the meson species indicates that the system does indeed reach chemical equilibrium. It is observed that the pions have a large amount of the particle number densities and the reason could be the decay in the heavier mesons and other particles produced in the system to form the pions. The saturation of the particle number densities indicate the realization of the local equilibrium, and thus as conclusion, the chemical equilibrium of the system has been reached as in both figures the saturation times are the same for all three mesons regardless the shape of each meson.

In figure 1 the rhos and kaons particles decreases to their equilibrium state while pions quickly reach equilibrium state and never decreases. In figure 2 where the energy density is high, a change of shapes is observed for these meson species where by the pions and rhos increase to equilibrium state while the kaon still decreases to equilibrium state. 

In figure 1 where $\varepsilon = 0.3$ GeV/fm$^{3}$ the equilibrium time for all meson species is around $t = 22 $ fm/c and for figure 2 at higher energy density of $\varepsilon = 0.4$ GeV/fm$^{3}$, the equilibrium time is observed to have increased to $t = 32 $ fm/c. These results show that increase in energy density influence the particle multiplicity inside the periodic box which affect the equilibration time. The higher the energy density the longer it takes for the mesons to reach saturation state due to the fact that the particles have higher energies and they can collide many times before they reach equilibrium state. 
for each meson reason being that particle multiplicity inside the periodic box is different for all three meson species. It was observed that when increase in energy density the saturation time decreases and the number density which is due to large particle multiplicity because of high energy inside the periodic box. 

%%\newpage

\subsection{Thermal Equilibrium and temperature}
%\quad
%\newline
In this subsection the thermal equilibrium and the temperature from the energy distributions of different light meson species are studied. The possibility of the thermal equilibrium of the hadronic matter is studied by examining the energy distribution of the system in a box with periodic boundary conditions using UrQMD model. The particle spectra is given by the momentum distribution as
\begin{equation}
\frac{dN_{i}}{d^{3}p}=\frac{dN}{4\pi EpdE} \propto Ce^{\left( -\beta E_{i}\right) }.
\end{equation}
Figure 3 and figure 4 represent the time evolution of energy spectra of different meson species. The linear lines are fitted using the Boltzmann distribution which is aproximated by $C\exp(-\beta E_{i})$ from Eq. 5 where $\beta = 1/T$ is the slope parameter of the distribution and $E_{i}$ is the energy of particle $i$.

Figure 3 and figure 4 represent the energy spectra of meson species namely the pion, the kaon and the rho particles produced during the ultra-relativistic heavy ion collisions simulated using the UrQMD model in a box. The results are plotted as a function of kinetic energy $K = E - m$, so that the horizontal axes for all the particle species coincide \cite{sasaki2001study}. 
The figures where produced at different energy densities respectively, with $\varepsilon$ = 0.2 GeV/fm$^{3}$ for figure 3 and $\varepsilon$ = 0.3 GeV/fm$^{3}$ for figure 4. From 
the figures 3 and 4 it is observed that the slopes of the energy distribution converges to common values of temperatures at different times above $t$ = 180 fm/c for $\varepsilon$ = 0.2 GeV/fm$^{3}$ and above $t$ = 250 fm/c for $\varepsilon$ = 0.3 GeV/fm$^{3}$. In thermal equilibrium the system is characterized by unique temeperature $T$ \cite{sasaki2001study}. The thermal temperatures where extracted from the equilibriumn state using the Boltzmann distribution $T$ = 118.3 MeV for $\varepsilon$ = 0.2 GeV/fm$^{3}$ and $T$ = 150.1 MeV for  $\varepsilon$ = 0.3 GeV/fm$^{3}$. These temperatures extracted from different energy densities from $\varepsilon$ = 0.1 - 0.5 GeV/fm$^{3}$ will be used to study EoS and thermal conductivity of hadronic matter.
 
%\newpage
\begin{figure}[H]
\includegraphics[width=0.35\textwidth]{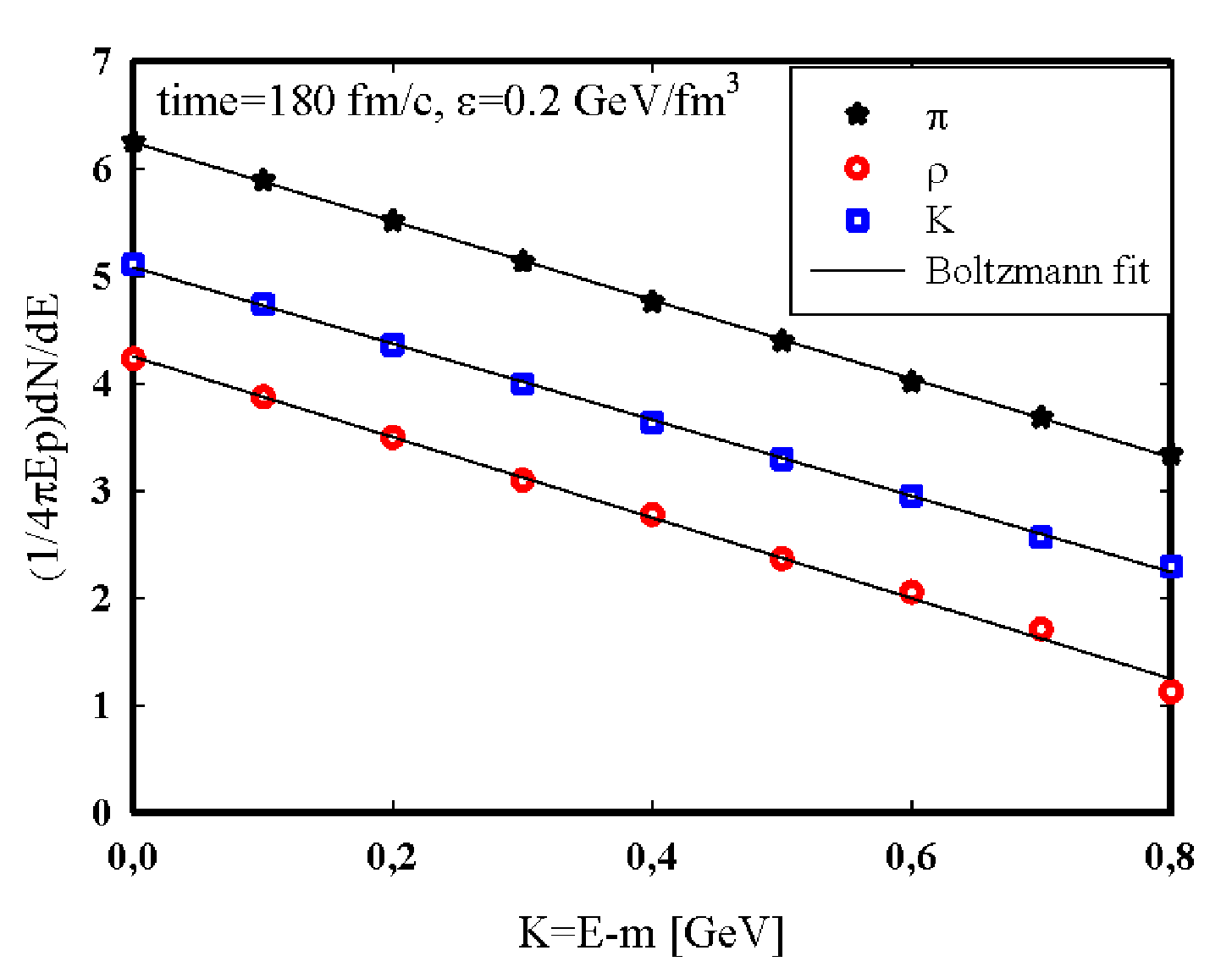}
\caption{ The energy distributions of the $\pi$, the $\rho$ and the $K$ at $\varepsilon$ = 0.2 GeV/fm$^{3}$ and $t$ = 180 fm/c. The lines are the Boltzmann fit which gives the extracted temperature of $T$ = 118.3 MeV.}
\label{energy}
\end{figure}
%\newpage

\begin{figure}[H]
\includegraphics[width=0.35\textwidth]{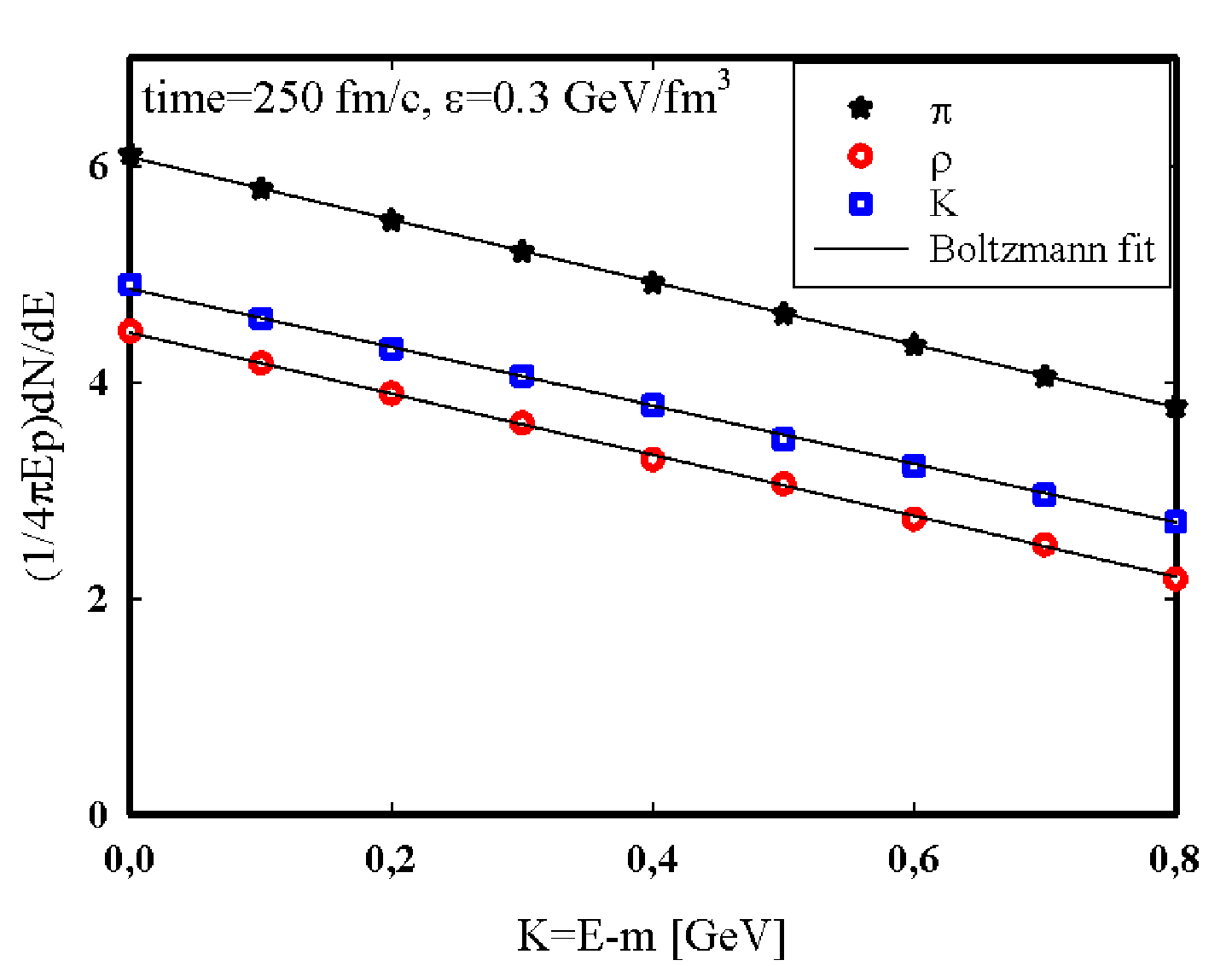}
\caption{ The energy distributions of the $\pi$, the $\rho$ and the $K$ at $\varepsilon$ = 0.3 GeV/fm$^{3}$ and $t$ = 250 fm/c. The lines are the Boltzmann fit which gives the extracted temperature of $T$ = 150.1 MeV.}
\label{spectra}
\end{figure}
 
\section{Hadronic gas model (EoS)}
%\quad
%\newline
The hadron abundances and the ratios have been suggested as the possible signatures for the exotic states and the phase transitions of the nuclear matter \cite{bleicher1999distinguishing}. The equation of state of hot and dense hadron gas provides the valuable information regarding the nature of the hadron gas. These signatures have been applied to the study of chemical equilibration in the relativistic heavy ion reactions. The properties like the temperatures, the entropies and chemical potentials of the hadronic matter have been extracted assuming thermal and chemical equilibrium \cite{bleicher1999relativistic}. 

In this section the EoS for a hadron gas is studied from the UrQMD calculations. The equation of state can be studied using a statistical model which is described by the grand canonical ensemble of non-interacting hadrons in an equilibrium sate at temperature $T$ as presented in \cite{muronga2004shear,bass1998microscopic}. A large number of studies have been done to study EoS of hadron gas \cite{muronga2004shear,bass1998microscopic,belkacem1998equation,pal2010shear}. For this study the focus is only on the EoS of hadrong gas made out of  the $\pi$, the $\rho$ and the $K$ calculated from the UrQMD model behaves. This is done through studying the evolution of pressure and energy density with temperature.

Some of the thermodynamic quantities used to calculate EoS are the energy density given by

\begin{equation}
\varepsilon=\frac{1}{V}\sum_{i=1}^{all-particles} E_{i} \, ,
\end{equation}
and the pressure which is given by

\begin{equation}
P=\frac{1}{3V}\sum_{i=1}^{all-particles}\frac{p^{2}_{i}}{E_{i}} .
\end{equation}

Figure 5 and Figure 6 represent the EoS of hadronic matter namely the pressure and energy density as function of temperature. The energy density used here is the same energy density used as an input parameter during the initialization of the ultra-relativistic heavy ion collision in the UrQMD. The pressure is then calculated from the collision results using Eq. (7).
 
The thermal temperatures used here are extracted from the thermal equilibrium (section 3.2) using the Boltzmann distribution function given as $C\exp({\frac{E}{T}})$.
%\newpage
\begin{figure}[H]
\includegraphics[width=0.35\textwidth]{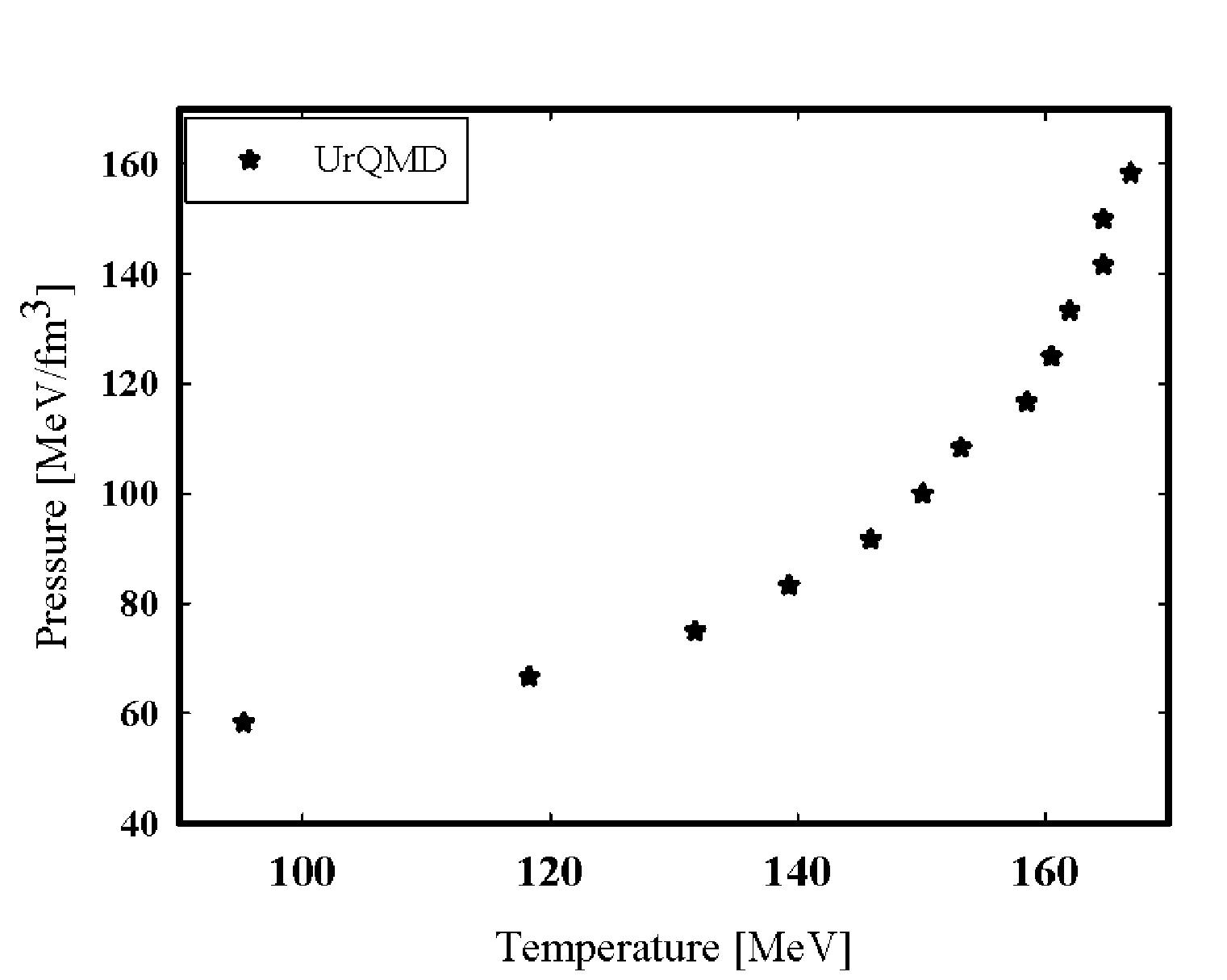}
\caption{The equation of state of a mixed hadron gas of the $\pi$, the $\rho$ and the $K$ at a finite temperature (100 MeV $<$ $T$ $<$ 200
MeV). The pressure of the hadronic gas is plotted as a function of a temperature.}
\label{equastion}
\end{figure}
%\newpage
\begin{figure}[H]
\includegraphics[width=0.35\textwidth]{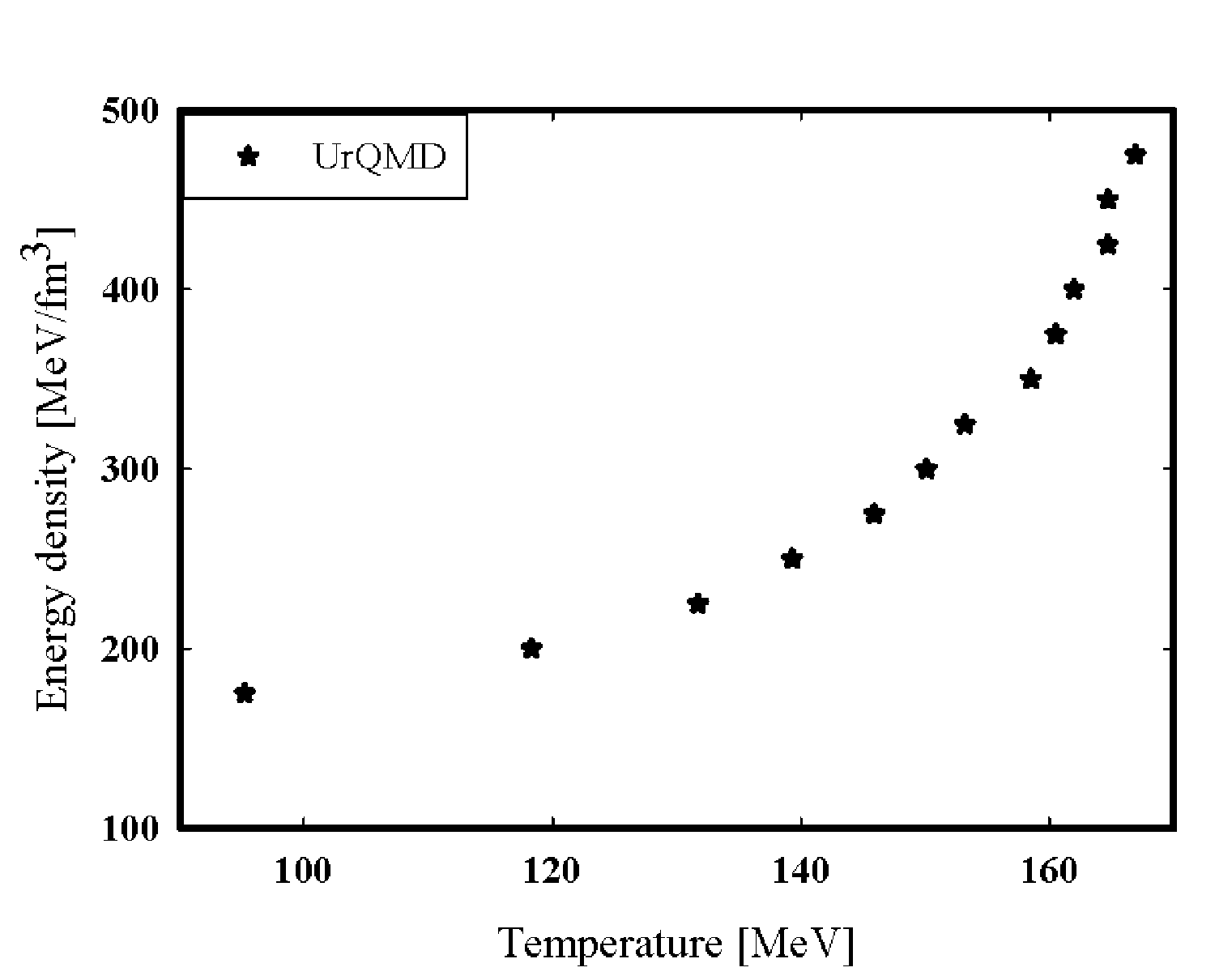}
\caption{The equation of state of a mixed hadron gas of the $\pi$, the $\rho$ and the $K$ at finite temperature (100 MeV $<$ $T$ $<$ 200
MeV). The energy density of hadronic gas is plotted as a function of temperature.}
\label{state}
\end{figure}
%\newpage

From the above figure 5 and figure 6, both pressure and energy density increases with an increase in the temperature. The results are in good comparison with those sustained in \cite{muronga2004shear,bass1998microscopic}. 
%The fitted line in both figure 5 and figure 6 represents the power law fit.

The focus is on the hadronic scale temperature of (100 MeV $<$ $T$ $<$ 200
MeV) and the zero baryon number density which is expected to be realized in the central high energy nuclear collisions \cite{muronga2004shear}. The pressure and energy density grows with temperature and starts to saturate just after $T$ = 150 MeV which might indicates that there is a change in phase transition of the hadron gas. In figure 5 and figure 6 at low temperatures between $T$ = 90 MeV to $T$ = 150 MeV the power law $T^{2}$ of hadron gas behaves differently than the power law $T^{4}$ at high temperature between $T$ = 155 MeV to $T$ = 170 MeV. The power law of the solid line fit is $T^{4}$ which is the same as that of hadron gas at high temperatures.
 
This results concludes that one can only calculate the thermal conductivity of this system between 100 MeV and 160 MeV as above 160 MeV while the temperature increases the system start to behave like that of massless particles called QGP. The above results in figure 5 and figure 6 shows that the pressure and the energy density increases to saturation state with temperature which indicate phase transion to a new state of matter the QGP. 

\section{Thermal conductivity coefficient}
%%\quad
%\newline
The transport coefficients such as the thermal conductivity $\kappa$ and the shear viscosity $\eta$ characterize the dynamics of the fluctuations of the dissipative fluxes in a medium \cite{muronga2004shear}. 
 
The most often used method to investigate these coefficient is either through employing the kinetic theory or the field theory using the Green-Kubo formula \cite{muronga2004shear}. 
 
The Green-Kubo relations for calculation of transport coefficients of shear viscosity, thermal conductivity, thermal diffusion and mutual diffusion for a binary mixture of hard spheres as well as for the calculation of diffusion coefficient of a hadron \cite{demir2014hadronic,muronga2004shear,wesp2011calculation,bocquet1994hydrodynamic}. 

A knowledge of various transport coefficients is important for the dissipative fluid dynamical model. One can calculate the coefficient of thermal conductivity from the fluctuation-dissipative theorem. The fluctuation-dissipative theorem tells us that the thermal conductivity is given by the heat current-current correlations \cite{muroya2007transport}.

The Green and Kubo showed that the transport coefficients like heat conductivity, shear and the bulk viscosities can be related to the correlation functions of the
corresponding flux or the tensor in the thermal equilibrium \cite{wesp2011calculation}. The Green-Kubo formalism relates linear transport coefficients to near-equilibrium correlations of dissipative fluxes and treats dissipative fluxes as perturbations to local thermal equilibrium \cite{demir2009extracting}. The relevant formular for Green-Kubo relation for thermal conductivity can be written as \cite{muronga2008generalized}

\begin{equation}
\kappa =\frac{V}{3T^{2}}\int_{0}^{\infty}\left\langle \mathbf{q}_{i}\left( 0\right). \mathbf{q}_{i}\left( t\right) \right\rangle dt,
\end{equation}

In Eq. (8) the brackets \textless...\textgreater\ stand for the equilibrium average, and no summation is implied over the repeated indices \cite{muronga2004shear,muronga2008generalized} and $\kappa$ is the thermal conductivity. 
%The time correlation function for the thermal conductivity is also given as \cite{wesp2011calculation,muronga2008generalized}  
% \begin{equation}
%\left\langle  \mathbf{q}_{i}\left( 0\right) \mathbf{q}_{i}\left( 0\right) \right\rangle = \kappa 3T^{2}\left( \tau_{q}V\right) ^{-1}\exp\left( -t/\tau_{q}\right) ,
%\end{equation}
The vector $\mathbf{q}_{i}$ is the Eckart's heat current along the $i$ = $x, y$ and $z$ axis which is defined as

\begin{equation}
\mathbf{q}^{i}=\frac{1}{V}\sum_{k=1}^{N} \mathbf{p}^{i} \left( \frac{\mathbf{p}^{2}}{\mathbf{p}_{0}^{2}}\right) ,
\end{equation}
where by $\mathbf{p}^{i}$ is the momentum along the $i$ = $x, y$ and $z$ axis and $\mathbf{p}^{2} = \mathbf{p}_{0}^{2} - \mathbf{p}_{x}^{2} - \mathbf{p}_{y}^{2} - \mathbf{p}_{z}^{2}$ which can be extracted from the UrQMD model output file and $\tau_{q}$ is the relaxation time of the heat current. If the evolution of the fluctuations of the fluxes is described by the Maxwell-Cattaneo 
\cite{muronga2008generalized}, after the integration, Eq. (8) is reduces to 

\begin{equation}
\kappa=\tau_{q}\frac{V}{T^{2}}\left\langle \mathbf{q}_{i}\left( 0\right) \mathbf{q}_{i}\left( 0\right) \right\rangle .
\end{equation}
We adopt a relativistic microscopic model, namely the UrQMD \cite{bass1998microscopic} and perform a molecular-dynamics for a hadronic gas of mesons in a box to compute the thermal conductivity coefficient of the hadron gas.

Figure 7 represents the square expectation of heat current, 
figure 8 represents thermal conductivity, figure 9 represents thermal conductivity times temperature and figure 10 represents the relaxation time for heat conductivity of a hadron gas as a function of temperature.

%\newpage
\begin{figure}[H]
\includegraphics[width=0.35\textwidth]{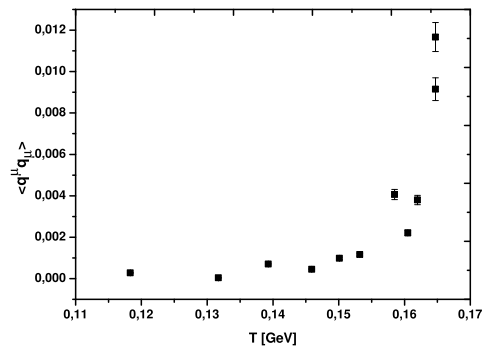}
\caption{The square expectation value of heat current for hadron gas of the $\pi$, the $\rho$ and the $K$ as a function of temperature.}
\label{squarex}
\end{figure}

\begin{figure}[H]
\includegraphics[width=0.35\textwidth]{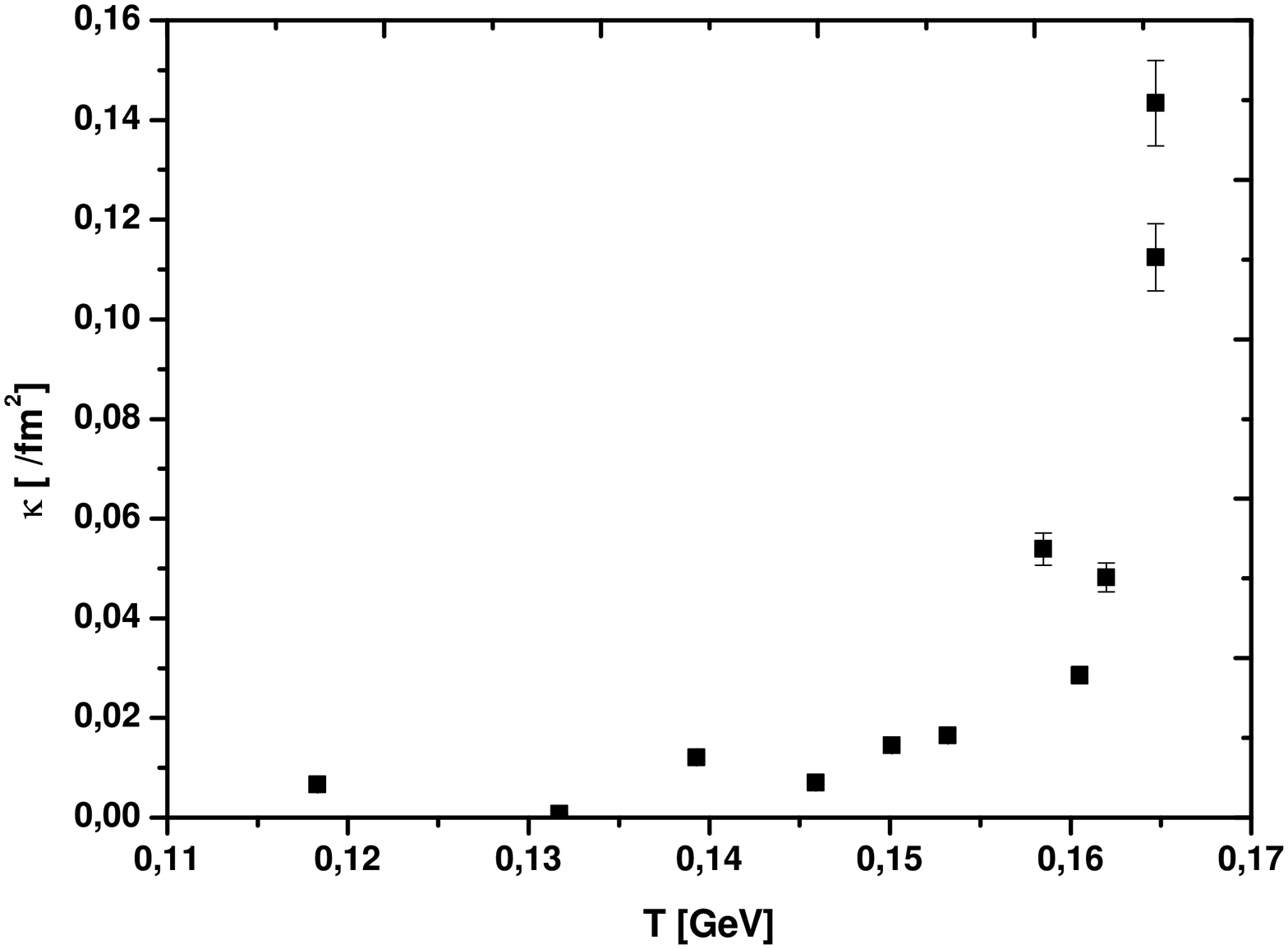}
\caption{The thermal conductivity of a hadron gas of the $\pi$, the $\rho$ and the $K$ as a function of temperature.}
\label{tconduct}
\end{figure}
%\newpage
%\newpage
\begin{figure}[H]
\includegraphics[width=0.35\textwidth]{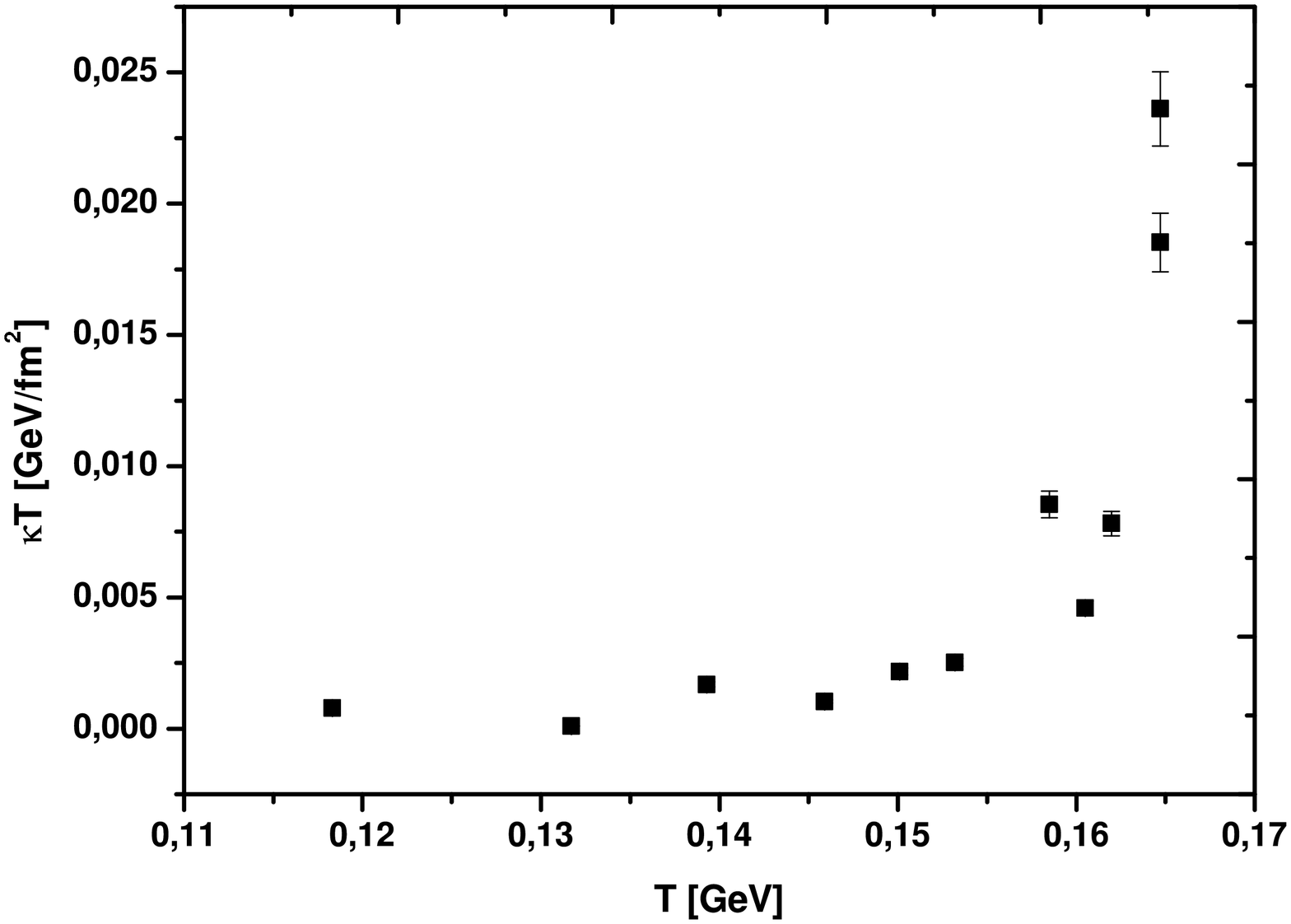}
\caption{The product of thermal conductivity and temperature of a hadron gas of the $\pi$, the $\rho$ and the $K$ as a function of temperature.}
\label{product}
\end{figure}

%\newpage
\begin{figure}[H]
\includegraphics[width=0.35\textwidth]{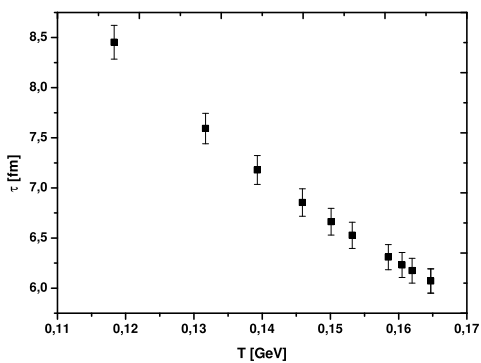}
\caption{The relaxation time for heat conductivity of a hadron gas of the $\pi$, the $\rho$ and the $K$ as a function of temperature.}
\label{tau}
\end{figure}

Figure 7 shows the square expectation of heat current results obtained from UrQMD model. The heat current increases with increase in temperature. These results are in good comparison with those obtained in \cite{muroya2007transport}, where a different model named URASIMA was used. The UrQMD square expectation of heat current is much smaller from that obtained in \cite{muroya2007transport}, reason might be that for this study we only considered a situation of only meson species and zero baryon number.

Figure 8 shows thermal conductivity of hadron gas and figure 9 shows the product of thermal conductivity coefficient times temperatureas a function of temperature calculated from UrQMD model. Both, $\kappa$, and, $\kappa T$, increases with an increase in temperature and the saturation is reached below $T$ = 0.17 GeV where the hadronic gas with zero baryon number density is expected to be realized in the central high energy nuclear collisions \cite{muronga2004shear}. The error bars are randomly plotted with the graphs.
 
According to our simulation we observed a strong temperature dependence. The temperature behavior of the product of the thermal conductivity and temperature is $\kappa T$ $\sim$ $T^{3.88}$ and for thermal conductvity is $\kappa$ $\sim$ $T^{2.88}$. The temperature dependence of thermal conductivity in this study is greater $T^{3.88} > T^{2}$ than the one reported by \cite{torres2012hadronic} where by the author used Effective Field Theory and it is less $T^{3.88} < T^{4}$ than  the one reported by \cite{muroya2007transport} whereby the author used different simulation model called URASIMA and different system which consist of baryon number density. 

Due to less number of studies done under thermal conductivity coefficient it is difficult to make a proper conclusion from the results obtained, but from the comparison with the little study done already which is related to this topic one can conclude that the results are in good comparison with those reported by \cite{muroya2007transport}. At the moment it is not very much clear where the large fluctuation around $T$ = 160 MeV comes from in the above figures. Thus a similar study will be done in future which will include the baryon number density and different mesons species at higher energies and large number of events in order to check if one of these factors does play a role for this large fluctuation. For this study the aim was just to test if it is possible to compute thermal conductivity transport coefficient from the UrQMD model using the Green-Kubo relations. The results in this study for figure 8 and figure 9 are smaller and that of figure 10 are larger than those reported in  \cite{muroya2007transport}. The reason might be that the author was using different input parameters for the initialization of the system. In our case we only consider a box of only meson and zero baryon number densities. 

Figure 10 shows the relaxation time for the heat flux of hot hadron gas as a function of temperature calculated from UrQMD model by fitting the heat correlation functions. The heat relaxation time decrease with an increase in temperature similarly to the one reported in \cite{muronga2004shear,muroya2007transport}.

%\newpage
\section{Conclusion}
%\newline
Thermal conductivity transport coefficient of hadron gas was studied using a microscopic transport model the UrQMD to simulate the ultra-relativistic heavy ion collisions in a box with a periodic boundary conditions and $V$. From the equilibrium state, thermal conductivity was calculated from the heat current - current correlations using the Green-Kubo relations.

It was observed that the square expectation of the heat current, thermal conductivity and the product of thermal conductivity and temperature increase with an increase in temperature. The saturation temperature for square expectation of the heat current, thermal conductivity and the product of thermal conductivity and temperature is below $T = 170$ MeV similarly to that of the equation of state. The power law shows a high temperature dependence for both $\kappa$ and $\kappa T$. The heat current relaxation time was calculated from heat correlation function. The relaxation time decrease with an increase in temperature. The results of heat current is in good comparison with that of  \cite{muroya2007transport} with regards to the behaviour of the graph.   

From the presented results, it can be concluded that it is possible to calculate thermal conductivity transport coefficient using the UrQMD model. More study is still required for better understanding of the results and the coefficient. The future studies will be to focus on how the thermal conductivity transport coefficient is affected by adding different number of meson species in the box including baryon number density in order to compare with other studies such as that reported by \cite{muroya2007transport,rougemont2015suppression}, as well as to compare to those who used different model and statistical approach \cite{torres2012hadronic,greif2013heat}.

\begin{acknowledgments}
I would like to thank everyone who helped me with this paper in terms of corrections and discussions. Mr Snymann Andrea who helped me with computational software's needed for this studies. Financial support from NITheP is acknowledged.

\end{acknowledgments}

\renewcommand{\bibname}{References}
\nocite{*}
\bibliographystyle{iopart-num}
\bibliography{apssamp}% Produces the bibliography via BibTeX.

\providecommand{\newblock}{}
\begin{thebibliography}{10}
\expandafter\ifx\csname url\endcsname\relax
  \def\url#1{{\tt #1}}\fi
\expandafter\ifx\csname urlprefix\endcsname\relax\def\urlprefix{URL }\fi
\providecommand{\eprint}[2][]{\url{#2}}

\bibitem{steinheimer2013nonthermal}
Steinheimer J, Aichelin J and Bleicher M 2013 {\em Physical review letters\/}
  {\bf 110} 042501

\bibitem{campbell2011vector}
Campbell J~M, Ellis R~K and Williams C 2011 {\em Journal of High Energy
  Physics\/} {\bf 2011} 1--36

\bibitem{bratkovskaya2000aspects}
Bratkovskaya E, Cassing W, Greiner C, Effenberger M, Mosel U and Sibirtsev A
  2000 {\em Nuclear Physics A\/} {\bf 675} 661--691

\bibitem{demir2014hadronic}
Demir N and Wiranata A 2014 {\em Journal of Physics: Conference Series\/} vol
  535 (IOP Publishing) p 012018

\bibitem{muronga2004shear}
Muronga A 2004 {\em Physical Review C\/} {\bf 69} 044901

\bibitem{gorenstein2008viscosity}
Gorenstein M, Hauer M and Moroz O 2008 {\em Physical Review C\/} {\bf 77}
  024911

\bibitem{bass1998microscopic}
Bass S~A, Belkacem M, Bleicher M, Brandstetter M, Bravina L, Ernst C, Gerland
  L, Hofmann M, Hofmann S, Konopka J {\em et~al.\/} 1998 {\em Progress in
  Particle and Nuclear Physics\/} {\bf 41} 255--369

\bibitem{PhysRevC.79.044901}
Mitrovski M, Schuster T, Gr\"af G, Petersen H and Bleicher M 2009 {\em Phys.
  Rev. C\/} {\bf 79}(4) 044901
  \urlprefix\url{http://link.aps.org/doi/10.1103/PhysRevC.79.044901}

\bibitem{becattini2013hadron}
Becattini F, Bleicher M, Kollegger T, Schuster T, Steinheimer J and Stock R
  2013 {\em Physical review letters\/} {\bf 111} 082302

\bibitem{bleicher2015recent}
Bleicher M, Endres S, Steinheimer J and van Hees H 2015 {\em arXiv preprint
  arXiv:1503.07371\/}

\bibitem{song1997chemical}
Song C and Koch V 1997 {\em Physical Review C\/} {\bf 55} 3026

\bibitem{li2009effects}
Li Q, Steinheimer J, Petersen H, Bleicher M and St{\"o}cker H 2009 {\em Physics
  Letters B\/} {\bf 674} 111--116

\bibitem{muronga2007relativistic}
Muronga A 2007 {\em Physical Review C\/} {\bf 76} 014910

\bibitem{belkacem1998equation}
Belkacem M, Brandstetter M, Bass S~A, Bleicher M, Bravina L, Gorenstein M~I,
  Konopka J, Neise L, Spieles C, Soff S {\em et~al.\/} 1998 {\em Physical
  Review C\/} {\bf 58} 1727

\bibitem{bleicher1999relativistic}
Bleicher M, Zabrodin E, Spieles C, Bass S~A, Ernst C, Soff S, Bravina L,
  Belkacem M, Weber H, St{\"o}cker H {\em et~al.\/} 1999 {\em Journal of
  Physics G: Nuclear and Particle Physics\/} {\bf 25} 1859

\bibitem{panda2005strong}
collaboration P {\em et~al.\/} 2005 {\em Technical Progress Report, GSI,
  Darmstadt\/}

\bibitem{sasaki2001study}
Sasaki N 2001 {\em Progress of theoretical Physics\/} {\bf 106} 783--805

\bibitem{bleicher1999distinguishing}
Bleicher M, Reiter M, Dumitru A, Brachmann J, Spieles C, Bass S~A, St{\"o}cker
  H and Greiner W 1999 {\em Physical Review C\/} {\bf 59} R1844

\bibitem{pal2010shear}
Pal S 2010 {\em Physics Letters B\/} {\bf 684} 211--215

\bibitem{wesp2011calculation}
Wesp C, El A, Reining F, Xu Z, Bouras I and Greiner C 2011 {\em Physical Review
  C\/} {\bf 84} 054911

\bibitem{bocquet1994hydrodynamic}
Bocquet L and Barrat J~L 1994 {\em Physical review E\/} {\bf 49} 3079

\bibitem{muroya2007transport}
Muroya S 2007 {\em arXiv preprint hep-ph/0702220\/}

\bibitem{demir2009extracting}
Demir N and Bass S~A 2009 {\em The European Physical Journal C\/} {\bf 62}
  63--68

\bibitem{muronga2008generalized}
Muronga A 2008 {\em The European Physical Journal-Special Topics\/} {\bf 155}
  107--113

\bibitem{torres2012hadronic}
Torres-Rincon J~M 2012 {\em arXiv preprint arXiv:1205.0782\/}

\bibitem{rougemont2015suppression}
Rougemont R, Noronha J and Noronha-Hostler J 2015 {\em Physical Review
  Letters\/} {\bf 115} 202301

\bibitem{greif2013heat}
Greif M, Reining F, Bouras I, Denicol G, Xu Z and Greiner C 2013 {\em Physical
  Review E\/} {\bf 87} 033019

\bibitem{groot}
de~Groot S, van Leeuwen W and van Weert C 1980 {\em Relativistic Kinetic
  Theory\/} (North-Holland Publishing Company Amsterdam.New York.Oxford) ISBN
  0-444-85453-3

\bibitem{baier2008relativistic}
Baier R, Romatschke P, Son D~T, Starinets A~O and Stephanov M~A 2008 {\em
  Journal of High Energy Physics\/} {\bf 2008} 100

\bibitem{luzum2008conformal}
Luzum M and Romatschke P 2008 {\em Physical Review C\/} {\bf 78} 034915

\bibitem{romatschke2010new}
Romatschke P 2010 {\em International Journal of Modern Physics E\/} {\bf 19}
  1--53

\bibitem{bouras2010investigation}
Bouras I, Molnar E, Niemi H, Xu Z, El A, Fochler O, Greiner C and Rischke D
  2010 {\em Physical Review C\/} {\bf 82} 024910

\bibitem{heinz2006dissipative}
Heinz U, Song H and Chaudhuri A~K 2006 {\em Physical Review C\/} {\bf 73}
  034904

\bibitem{jeon1996quantum}
Jeon S and Yaffe L~G 1996 {\em Physical Review D\/} {\bf 53} 5799

\bibitem{csernai1994introduction}
Csernai L~P 1994 {\em Introduction to relativistic heavy ion collisions\/}
  (Wiley New York)

\end{thebibliography}
\end{document}